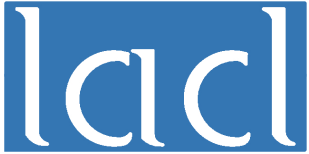 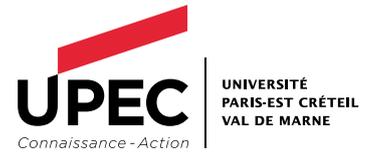

# Generic Environments in Coq

**Emmanuel Polonowski**

Laboratoire d'Algorithmique, Complexité et Logique (LACL)
Département d'Informatique
Université Paris-Est Créteil – Val de Marne, Faculté des Sciences et Technologie
61, Avenue du Général de Gaulle, 94010 Créteil cedex, France
Tel.: (33)(1) 45 17 16 47, Fax: (33)(1) 45 17 66 01

Laboratory of Algorithmics, Complexity and Logic (LACL)

University Paris-Est Créteil

Technical Report **TR–LACL–2011–3**

Emmanuel Polonowski.

*Generic Environments in Coq*

# Generic Environments in Coq

by Emmanuel Polonowski

LACL, University Paris-East Créteil

*Email:* emmanuel.polonowski@u-pec.fr

**Abstract**

We introduce a library which provides an abstract data type of environments, as a functor parameterized by a module defining variables, and a function which builds environments for such variables with any Type of type. Usual operations over environments are defined, along with an extensive set of basic and more advanced properties. Moreover, we give an implementation using lists satisfying all the required properties.

## 1 Introduction

A standard usage of proof assistants is to define type systems for a wide range of formal languages, and to prove properties about them. To achieve this goal, one has to work around the notion of typing environment, from the definition of the type system up to the core of the proofs. The definition of such typing environments is mainly a programming activity, but it often requires to prove a lot of small uninteresting (and easy to prove) lemmas that eventually are packed in some auxiliary file, unordered and probably redundant w.r.t. themselves.

Among his libraries for the locally nameless infrastructure (see [1]), Arthur Charguéraud proposed some years ago a generic module for environments, that contained a function which builds environments for a given type, specialized for a previously defined notion of variable. This library was an elegant way to deal (almost) abstractly with environments.

The work we present here is a complete rewriting of this library of environments, reorganized and augmented in several ways. Firstly, it has be written in Coq v8.3 without any use of additional library or tactic; we believe it to be more stable than its previous version. Secondly, instead of importing a module with the definition of variables, we define our library as a functor parameterized with such a module, requiring only the decidability of equality over the variables. Thirdly, the set of basic operations has been extended with : a remove operation for a single variable or a list of variables, and an update operation which changes the types associated to variables accordingly to another environment. Fourthly, we added an equivalence relation over environments, based upon environment bindings inclusion, which can be used to work on the theory of typing environments. Fifthly, as we follow Charguéraud in providing an implementation of environments with lists, we enforce the abstract-data-type discipline with the help of the module system of Coq.

We tried to be as exhaustive as possible in the choice of the provided properties, and with the least possible redundancy. We believe that the set of around 250 basic and more advanced properties might be enough for any use of this library. As a final and validating property, we prove the equivalence between updating an environment with a binding and adding it after having removed the previous binding. Even if the result is of few theoretical interest, it ensures that the definitions are sufficient to prove arbitrary abstract properties over environments.

The paper is organized as follows: Section 2 briefly describes the basic definitions of our environments, Section 3 exhibits several standard properties of environments and the proof of equivalence between updating and removal followed by concatenation, Section 4 concludes with a discussion about environments and choices of implementation.

Source code and automatically generated html documentation are of course available [3].

# 2 Defining a generic notion of environment

## 2.1 Structure of the library

The structure of the library is given by three library files organized as follows.

Firstly, the module of variables, that has to be implemented and then given as parameter to the generic environment functor.

```
Module Type Generic_Var.
  Parameter TVar : Type.
  Parameter eq_var_dec : forall x y : TVar, {x = y} + {x <> y}.
End Generic_Var.
```

As one can see, its definition requires only the decidability of equality, but it could be augmented with additional definitions if it were to be used, for instance, to define a module which deals with freshness.

Secondly, the functor of abstract generic environments, which is parameterized by the module of variables.

```
Module Type Generic_Env_Type (Var : Generic_Var).
  ...
End Generic_Env_Type.
```

Thirdly, the functor of generic environments implemented with lists and the opaque exported functor.

```
Module Generic_Env_List_Def (Var : Generic_Var).
  ...
End Generic_Env_List_Def.

Module Generic_Env_List (Var : Generic_Var) : Generic_Env_Type (Var) :=
    Generic_Env_List_Def Var.
```

The module `Generic_Env_List` is the opaque functor to be used by the library users.

## 2.2 Basic definitions of environments

We present here the basic definitions of environments.

The main function of the module is the following:

```
Parameter gen_env : Type -> Type.
```

It takes as a parameter the type of types that are to be associated with variables in the returned environment.

Two operations are defined to build environments with a single binding and several bindings given as a list of variable and a list of type:

```
Parameter single : TVar -> A -> gen_env A.
Parameter singles : list TVar -> list A -> gen_env A.
```

The basic operations on environments are the concatenation of two environments, the retrieving of a type associated to a variable that might belongs to it (we use the option type), the domain and the image of the environment seen as a sort of mapping:

```
Parameter concat : gen_env A -> gen_env A -> gen_env A.
Parameter get : TVar -> gen_env A -> option A.
Parameter dom : gen_env A -> list TVar.
```

```
Parameter img : gen_env A -> list A.
```

We provide as a basic operation the updating of an environment with the binding of another given a parameter. More explicitly, when updating an environment Γ with Δ, the resulting environment binds as Γ for the variables that are not in Δ, and as Δ for the variables of Γ that appears in Δ.

```
Parameter update : gen_env A -> gen_env A -> gen_env A.
```

At last, we provide two operations to remove bindings by their variables:

```
Parameter remove : TVar -> gen_env A -> gen_env A.
Parameter all_remove : list TVar -> gen_env A -> gen_env A.
```

The basic set of predicates over environments includes the tests of membership:

```
Parameter belongs : TVar -> gen_env A -> Prop.
Parameter all_belongs : list TVar -> gen_env A -> Prop.
Parameter notin : TVar -> gen_env A -> Prop.
Parameter all_notin : list TVar -> gen_env A -> Prop.
```

Of course, the main predicate is the binding assertion, to be used in the typing rules of a type system:

```
Definition binds (x : TVar) (v : A) (E : gen_env A) := get x E = Some v.
```

The extension of binds to several associations give rise to a notion of environment inclusion, defined as follows:

```
Definition all_binds (E F : gen_env A) := forall x v, binds x v E -> binds x v F.
```

We added a predicate of environment equivalence, defined as their mutual inclusion:

```
Definition eq (E F : gen_env A) := all_binds E F /\ all_binds F E.
```

Up to now, we deal with environments that possibly contain several bindings for the same variable. The following predicate assert the non-duplication of variables (where & is the concatenation of environments):

```
Inductive ok : gen_env A -> Prop :=
| ok_nil : ok (@empty A)
| ok_cons : forall x v F, ok F /\ x ∉ F -> ok (F & (x : v)).
```

Environments that are ok will have more properties, as the the commutativity of concatenation for instance.

## 3 Some properties of generic environments

Among the wide set of properties of environments, we find structural properties which helps to reason about the definitions, basic properties useful for more complex properties and external proofs, and advanced properties that might not be of any external use, but which provides intuitions about the robustness of our definitions. Here follows several properties organized in that order.

### 3.1 Structural properties

The simplest structural properties are those stating the neutrality of empty environments w.r.t. concatenation, and the associativity of the concatenation:

```
Parameter concat_empty_r : forall E, E & (@empty A) = E.
Parameter concat_empty_l : forall E, (@empty A) & E = E.
Parameter concat_assoc : forall E F G, E & (F & G) = (E & F) & G.
```

Properties about the construction of an environment from a list of variable and a list of type helps to catch the minimal requirements of the implementation:

```
Parameter singles_empty : nil :: nil = (@empty A).
Parameter singles_cons : forall x xs v vs,
  (x :: xs) :: (v :: vs) = (xs :: vs) & (x : v).
```

The `ok` predicate presented above was taken from the original implementation of Charguéraud. However, an interesting (almost) structural property is the equivalence between that predicate and the non-duplication of the domain of the environment, stated as follows:

```
Parameter ok_NoDup_dom_eq : forall E, ok E <-> List.NoDup (dom E).
```

## 3.2 Basic properties

We find also more complex properties, along with their inverse, that are mainly to be used in external proofs. For instance, the `ok` predicate for the concatenation of two environments:

```
Parameter ok_concat : forall E F,
  ok E -> ok F ->
  dom E ⊄ F -> dom F ⊄ E ->
  ok (E & F).
Parameter ok_concat_inv : forall E F,
  ok (E & F) ->
  ok E /\ ok F /\ dom E ⊄ F /\ dom F ⊄ E.
```

A lot of properties talks about the interactions between operations and predicates. For instance, the following property states that when we perform the updating of an environment with another which contains only disjoint bindings, then nothing happens:

```
Parameter update_notin : forall E F,
  (dom F) ⊄ E -> (E ::= F) = E.
```

The most important properties about environments are certainly those talking about bindings. The following one are very standard:

```
Parameter binds_eq_inv : forall x v w E, x : v ∈ E -> x : w ∈ E -> v = w.
Parameter binds_concat_r : forall x v F G, x : v ∈ G -> x : v ∈ (F & G).
Parameter binds_belongs : forall x v F, x : v ∈ F -> x ∈ F.
```

The inverse property of the last one is also quite useful:

```
Parameter belongs_binds : forall x F, x ∈ F -> exists v, x : v ∈ F.
```

We provide two decidability properties of the binds predicate. The first one for the existence of a binding for a given variable:

```
Parameter binds_dec_exists : forall x E,
  { v | x : v ∈ E } + { forall v, ¬ x : v ∈ E }.
```

The second one is the more usual decidability of the predicate. However, it requires the decidability of the equality over types in order to be provable:

```
Parameter binds_dec : forall x v E,
  (forall w w', { w = w' } + { ¬ w = w' }) ->
  { x : v ∈ E } + { ¬ x : v ∈ E }.
```

### 3.3 Advanced properties

When dealing with binding, environment inclusion and equivalence, we get to more complex properties. Here follows the decidability of the existence of a list of types associated with a given list of variables in an environment (where E ⊑ F is the environment inclusion):

```
Parameter all_binds_dec_exists : forall xs F,
  List.NoDup xs ->
  { vs | length xs = length vs /\ (xs :: vs) ⊑ F }
  + { forall vs, length xs = length vs -> ¬ (xs :: vs) ⊑ F }.
```

Environment inclusion is an order, as the following properties says (where ≍ is the equivalence over environments):

```
Parameter all_binds_refl : forall E F, E = F -> E ⊑ F.
Parameter all_binds_anti_sym : forall E F, E ⊑ F -> F ⊑ E -> E ≍ F.
Parameter all_binds_trans : forall E F G, E ⊑ F -> F ⊑ G -> E ⊑ G.
```

Moreover, we require that equivalence over environments satisfies the usual properties:

```
Parameter eq_refl : forall E F, E = F -> E ≍ F.
Parameter eq_sym : forall E F, E ≍ F -> F ≍ E.
Parameter eq_trans : forall E F G, E ≍ F -> F ≍ G -> E ≍ G.
```

With this equivalence, we can state interesting abstract properties, such as the commutativity of concatenation when environments are `ok`:

```
Parameter eq_concat_comm : forall E F, ok (E & F) -> (E & F) ≍ (F & E).
```

As a final example, we try to reason about operations on environments at this suitable abstract level. The following property assert the equivalence of updating and removal plus concatenation:

```
Parameter update_is_remove_concat : forall x v E,
  x ∈ E -> (E ::= (x : v)) ≍ ((E \ {x}) & (x : v)).
```

## 4 Discussion

**Should environments be built upon association lists or maps ?**

This is a natural question since environments are usually implemented as association lists. It is true that this library could be split in two parts: a library for association lists on one hand, and environments built upon it on the second hand. The critical point is about the duplication of variables: maps usually preserves the non-duplication of keys, while association lists are insensible about it. Here the choice is to allow duplication, discarding the use of Coq library for maps.

It will be interesting to develop a library for association lists, with enough basic properties to lighten the environment library.

**Should environments assume that no variable is duplicated ?**

Traditionally, no assumption is made on the duplication of variables in a typing environment, either discarding the question at the meta-level, or defining environment as stacks (or even both, using it as a stack without saying it). Indeed, the stack discipline works fine for environments even in presence of α-equivalence, since the typing relation is built structurally to respect the scope of binders. Our generic environment clearly enforces the stack discipline w.r.t. the concatenation to the right, stated informally by several properties (mainly `binds_concat_r` vs. `binds_concat_l`).

Could we state the non-duplication as a parameter of environments, and use only that kind of environments in every type systems ? The answer is no. Some languages are strongly built upon the stack discipline to encode different notions of binding, as in [2], and need to preserve all the occurrences of the same variable which are clearly distinct one from each other. There are other formal languages that are built upon slightly different notions of binding, and they would probably suffer for a lack of flexibility a notion of environment enforcing the non-duplication.

# Table of contents